\documentstyle[preprint,aps,epsf,graphicx]{revtex}
\begin{document}

%\tightenlines

\title{
$\Lambda\Lambda$ bond energy from the Nijmegen potentials }

\author{I. Vida\~na\footnote{Present address: Gesellschaft f\"{u}r Schwerionenforschung (GSI), Planckstr. 1, 
D-64291 
Darmstadt, Germany}}
\address{Dipartimento di Fisica ``Enrico Fermi'', Universit\`a di Pisa and INFN
Sezione di Pisa,
Via Buonarroti 2, I-56127 Pisa, Italy}
\author{A. Ramos and A. Polls}
\address{Departament d'Estructura i Constituents de la Mat\`eria,
Universitat de Barcelona, \\
Diagonal 647, E-08028 Barcelona, Spain
}

\date{\today}

\maketitle

\begin{abstract}

The $\Lambda\Lambda$ bond energy $\Delta B_{\Lambda \Lambda}$ in
$\Lambda\Lambda$ hypernuclei is obtained from a $G$-matrix
calculation which includes the coupling between the
$\Lambda\Lambda$, $\Xi N$ and $\Sigma\Sigma$ channels, as well as
the effect of Pauli blocking to all orders. The Nijmegen NSC97e
model is used as bare baryon-baryon interaction in the strangeness $S=-2$ sector. The
$\Lambda\Lambda$-$\Xi N$ coupling increases substantially the bond
energy with respect to the uncoupled $\Lambda\Lambda$ case. However, the additional
incorporation of the $\Sigma \Sigma$ channel, which couples
simultaneously to $\Lambda\Lambda$ and $\Xi N$ states, has a
surprisingly drastic effect and reduces the bond energy down to a
value closer to that obtained in an uncoupled calculation. We find
that a complete treatment of Pauli blocking reduces the repulsive
effect on the bond energy to about half of what was claimed
before.

%\end{abstract}

%\vspace{0.25cm}


%\noindent {\it Keywords: hypernuclei }

\end{abstract}

%\maketitle

%%%%%%%%%%%%%%%%%%%%%%%%%%%%%%%%%%%%%%%%%%%%%%%%%%%%%%%%%%%%%%%%%%%%%%%%%%%%%%%%%%%%

\section{Introduction}
\label{sec:intro}

Double-strange $\Lambda$ hypernuclei are nowadays the best systems
to investigate the properties of the $S=-2$ baryon-baryon
interaction. Emulsion experiments and subsequent analysis
\cite{danysz,prowse,aoki,dover} have reported the formation of a
few $\Lambda\Lambda$ hypernuclei, $^{6}_{\Lambda \Lambda}$He,
$^{10}_{\Lambda \Lambda}$Be and $^{13}_{\Lambda \Lambda}$B. From
the resulting $\Lambda \Lambda$ binding energies, a quite large
$\Lambda \Lambda$ bond energy of around 4-5 MeV emerged, contrary
to expectations from SU(3) \cite{nij99}. A series of theoretical
works, based either on phenomenological $S=-2$ baryon-baryon
interactions or realistic ones, have studied the properties of
double-$\Lambda$ hypernuclei for more than 30 years
\cite{dalitz:1964,bodmer:1965,ali:1967,bodmer:1984,Bodmer:rf,yamamoto,Portilho:zh,Portilho:th,Adam:ba,lanskoy1,hiyama,Carr:1997ce,Marcos:1997as,lanskoy:1998,Caro:1998kb,Nemura:1999qp,yamada:2000,vidana:2001}.

The recent finding at KEK of a new $^{6}_{\Lambda\Lambda}$He
candidate having a $\Lambda\Lambda$ bond energy of around 1 MeV
\cite{Takahashi:nm} has injected a renewed interest on this field.
Unless new experiments for the other $\Lambda\Lambda$ hypernuclei
also give lower binding energies in the future, it is now an open
question to reconcile theoretically the weak attraction found in
$^{6}_{\Lambda\Lambda}$He with the stronger attraction in the
other two heavier systems. Although some progress has been made in
Ref.~\cite{Albertus:2001pb}, where both short- and long-range
correlations were simultaneously treated, further investigations
are needed to completely settle this question. Filikhin and Gal
\cite{Filikhin:2002wm,Filikhin:2001yh,Filikhin:2003yh} report Faddeev-Yakubovsky
calculations, complementary to those carried out in the earlier
work of Ref.~\cite{hiyama} but using the new Nijmegen
interactions, not finding a simultaneous description of the
$^{10}_{\Lambda \Lambda}$Be and the new $^{6}_{\Lambda \Lambda}$He
binding energies. However, the nucleon Pauli blocking effect
affecting, through the coupling to $\Xi N$ states, the
$\Lambda\Lambda$ interaction when the particles are embedded in
the nuclear medium has not been considered in most of the earlier nor in
these recent works
\cite{Albertus:2001pb,Filikhin:2002wm,Filikhin:2001yh}. As
discussed in detail in Ref.~\cite{Carr:1997ce}, Pauli blocking
reduces substantially the additional attraction to the
$\Lambda\Lambda$ binding energy induced by the $\Lambda\Lambda \to
\Xi N$ conversion. Recently, an attempt to incorporate the Pauli
suppression effect has been made in Ref.~\cite{Myint:2002dp}, 
where a second order Pauli correcting term is introduced in the
intermediate states following the $\Lambda\Lambda \to \Xi N$
transition. The interaction used in that work is a
two-channel $(\Lambda\Lambda, \Xi N)$ Gaussian model, 
which implicitly includes the $\Sigma\Sigma$ coupling not only in the effective 
$\Lambda\Lambda$ interaction but also in the $\Lambda\Lambda \to \Xi N$ transition.
In fact, the important role of the coupling to $\Sigma\Sigma$ states has
been recently pointed out in Ref.~\cite{Afnan:2003ty} and explicitly worked out for the Nijmegen interactions in the variational calculation of Ref.~\cite{yamada}.

The purpose of the present work is to present a careful analysis
of the role of coupled channels on the $S=-2$ baryon-baryon
interaction in the medium, treating Pauli blocking effects to all
orders in all possible transition channels. For practical
purposes, most of the recent works
\cite{Filikhin:2002wm,Filikhin:2001yh,Filikhin:2003yh,Myint:2002dp} have used
simple parameterizations of the new Nijmegen potentials in terms
of the sum of a few gaussians. In contrast, we start from the
original Nijmegen model NSC97e\cite{nij99}, as done also in Ref.~\cite{yamada}. In our approach, we
solve the coupled-channel equation for the G-matrix in infinite
nuclear matter, and derive from it the $\Lambda\Lambda$ bond
energy in finite nuclei. With respect to existing calculations our
treatment of the finite system is very simple. This has the
practical advantage of permitting us to explore in depth the
different effects determining the $\Lambda\Lambda$ bond energy,
such as coupled channels or Pauli blocking to all orders. 
%We will
%show that there are substantial differences with respect to the
%approximate treatments found in the literature.

%%%%%%%%%%%%%%%%%%%%%%%%%%%%%%%%%%%%%%%%%%%%%%%%%%%%%%%%%%%%%%%%%%%%%%%%%%%%%%%%%%%%

\section{Formalism}
\label{sec:form}

The $\Lambda\Lambda$ bond energy $\Delta B_{\Lambda\Lambda}$ in $\Lambda\Lambda$ hypernuclei
is determined experimentally from the measurement of the binding energies of double- and
single-$\Lambda$ hypernuclei as
\begin{equation}
\Delta B_{\Lambda\Lambda}(^{A}_{\Lambda\Lambda}Z)= B_{\Lambda\Lambda}(^{A}_{\Lambda\Lambda}Z)
-2 B_{\Lambda}(^{A-1}_{\Lambda}Z) \ .
\label{eq:delta1}
\end{equation}
A reasonable estimation of this quantity when rearrangement
effects are small can be obtained from the value of the
$\Lambda\Lambda$ $G-$matrix element in a finite hypernucleus
\begin{equation}
\Delta B_{\Lambda\Lambda}(^{A}_{\Lambda\Lambda}Z) \approx -
\langle (0s_{1/2})_\Lambda (0s_{1/2})_\Lambda,J=0 \mid G \mid
(0s_{1/2})_\Lambda (0s_{1/2})_\Lambda,J=0  \rangle \ ,
\label{eq:delta3}
\end{equation}
where the two $\Lambda$ particles are assumed to be in the lowest
single particle state of an appropriate $\Lambda$-nucleus mean
field potential. We will compute the above matrix element from the
infinite nuclear matter one in the following way. First, we
construct the $\Lambda\Lambda$ $G-$matrix in infinite matter by
solving the well known Bethe--Goldstone equation which, in partial
wave decomposition and using the quantum numbers of the relative
and center--of--mass motion, reads
\begin{equation}
\begin{array}{c}
\langle Kq'L' S'J (\Lambda\Lambda) | G | KqLSJ (\Lambda\Lambda)\rangle =
\langle Kq'L' S'J (\Lambda\Lambda) | V | KqLSJ (\Lambda\Lambda)\rangle \nonumber  \\
\displaystyle{
+ \sum_{B_1B_2}\sum_{L''S''}\int dq''q''^2
\langle Kq'L' S'J (\Lambda\Lambda) | V | Kq''L''S''J (B_1B_2)\rangle} \nonumber \\
\times
\displaystyle{
\frac{Q_{B_1B_2}(K,q'')}{\Omega - \frac{K^2}{2(M_{B_1}+M_{B_2})} -
\frac{q''^2(M_{B_1}+M_{B_2})}{2M_{B_1}M_{B_2}} -
M_{B_1}-M_{B_2} +i\eta}} \nonumber \\
\times
\langle Kq''L'' S''J (B_1B_2) | G | KqLSJ (\Lambda\Lambda)\rangle  \ ,
\end{array}
\label{eq:gmat}
\end{equation}
where the labels $B_1B_2$ run over $\Lambda\Lambda, \Xi N$ and
$\Sigma\Sigma$ intermediate states. The starting energy $\Omega$
is taken equal to $2M_\Lambda-2B_{\Lambda}(^{A-1}_{\Lambda}Z)-
\Delta B_{\Lambda\Lambda}(^{A}_{\Lambda\Lambda}Z)=
2M_\Lambda-B_{\Lambda\Lambda}(^{A}_{\Lambda\Lambda}Z)$, where the experimental value of
$B_{\Lambda\Lambda}$ is taken for each hypernucleus, namely
7.25 MeV for 
$^{6}_{\Lambda\Lambda}$He \cite{Takahashi:nm}, 17.7 MeV for
$^{10}_{\Lambda\Lambda}$Be \cite{danysz} and 27.5 for
$^{13}_{\Lambda\Lambda}$B \cite{aoki}. In this way, we are considering the interaction of 
each $\Lambda$ particle not only with the nucleons in the nucleus but also
with the other $\Lambda$ particle.
The nuclear matter density to be used in the Pauli operator
$Q$ is determined, for each hypernucleus, as the average nuclear
density felt by the $\Lambda$ particle in that hypernucleus. This
is obtained by weighing the nuclear density at each point with the
probability of finding the $\Lambda$ particle:
\begin{equation}
\rho=\int \rho(r) \mid \Psi_{\Lambda}(r) \mid^2 d^3r \ ,
\label{eq:rho}
\end{equation}
where $\rho(r)$ is the nuclear density profile which is conveniently parameterized as
\begin{equation}
\rho(r)=\frac{\rho_0}{1+exp(\frac{r-R}{a})}
\label{eq:rho2}
\end{equation}
being
\begin{equation}
\rho_0=\frac{3}{4\pi}\frac{A_N}{R^3}\left( 1+\left( \frac{\pi a}{R}
\right)^2 \right) \ ,
\label{eq:rho0}
\end{equation}
with $a=0.52$ fm, $R=1.12{A_N}^{1/3}-0.86{A_N}^{-1/3}$ fm and
$A_N$ the number of nucleons in the hypernucleus. The
$\Lambda$ wave function is obtained by solving the Schr\"{o}dinger
equation using a Woods-Saxon $\Lambda$-nucleus potential with
parameters $(V_\Lambda,a_\Lambda,R_\Lambda)$ adjusted to reproduce the experimental
binding energy of the $\Lambda$ in the single-$\Lambda$
hypernucleus. For practical computational purposes, from the
resulting $\Lambda$ r.m.s. radius we derive the oscillator
parameter, $b_\Lambda$, of an equivalent harmonic oscillator wave
function which will then be used in obtaining the finite
hypernucleus two-body G-matrix elements of Eq.~(\ref{eq:delta3})
from the nuclear matter ones displayed in Eq.~(\ref{eq:gmat}).

In the next step, we express the two-body ket state
$|(0s_{1/2})_\Lambda (0s_{1/2})_\Lambda,J=0 \rangle$, built from the
$0s_{1/2}$ states of the equivalent harmonic oscillator
potential, in terms of momentum and angular variables
$|(k_1,l_1,j_1)_\Lambda(k_2,l_2,j_2)_\Lambda, J=0 \rangle$ in the
laboratory frame using
\begin{equation}
|(0s_{1/2})_\Lambda (0s_{1/2})_\Lambda,J=0 \rangle =
\int \int dk_1dk_2k_1^2k_2^2R_{00}(b_\Lambda k_1)R_{00}(b_\Lambda
k_2)|(k_1,0,\frac{1}{2})_\Lambda(k_2,0,\frac{1}{2})_\Lambda,J=0
\rangle \ ,
\label{eq:ho1}
\end{equation}
where $R_{nl}(x)$ is the corresponding harmonic oscillator function.

Finally, we express the two-body state with laboratory coordinates
in terms of the states with variables in the relative and
center-of-mass system, $|K q L S J (\Lambda \Lambda) \rangle$,
used in the solution of the Bethe--Goldstone equation
\begin{equation}
| (k_1,0,\frac{1}{2})_\Lambda(k_2,0,\frac{1}{2})_\Lambda,J=0 \rangle =
\int dKK^{2}\int dqq^{2} \langle Kq000 (\Lambda\Lambda) |
k_10\frac{1}{2}k_20\frac{1}{2},J=0\rangle | Kq000(\Lambda\Lambda)
\rangle \ , \label{eq:ho2}
\end{equation}
where $\langle Kq000 (\Lambda\Lambda)|
k_10\frac{1}{2}k_20\frac{1}{2},J=0 \rangle$ are the appropriate
transformation coefficients \cite{wo72,ku79} from the relative and
center-of-mass frame to the laboratory system. We note that the
only contribution comes from the partial wave $^1S_0$.
Transforming the bra state $ \langle(0s_{1/2})_\Lambda
(0s_{1/2})_\Lambda |$ in a similar way, one can finally evaluate
the $\Lambda\Lambda$ bond energy of Eq.~(\ref{eq:delta3}) in terms
of the infinite nuclear matter $\Lambda\Lambda$ $G-$matrix
elements.

In Table \ref{tab:tab1} we summarize all the parameters that allow
us to determine the relevant $\Lambda$ and nuclear properties
needed in the evaluation of the $\Lambda\Lambda$ bond energy for
the three hypernuclei studied in this work, $^{6}_{\Lambda
\Lambda}$He, $^{10}_{\Lambda \Lambda}$Be and $^{13}_{\Lambda
\Lambda}$B.

%%%%%%%%%%%%%%%%%%%%%%%%%%%%%%%%%%%%%%%%%%%%%%%%%%%%%%%%%%%%%%%%%%%%%%%%%%%%%%%%%%%%%%%%%%%%%%

\section{Results}
\label{sec:results}

\begin{figure}[htb]
\centerline{
     \includegraphics[width=\textwidth]{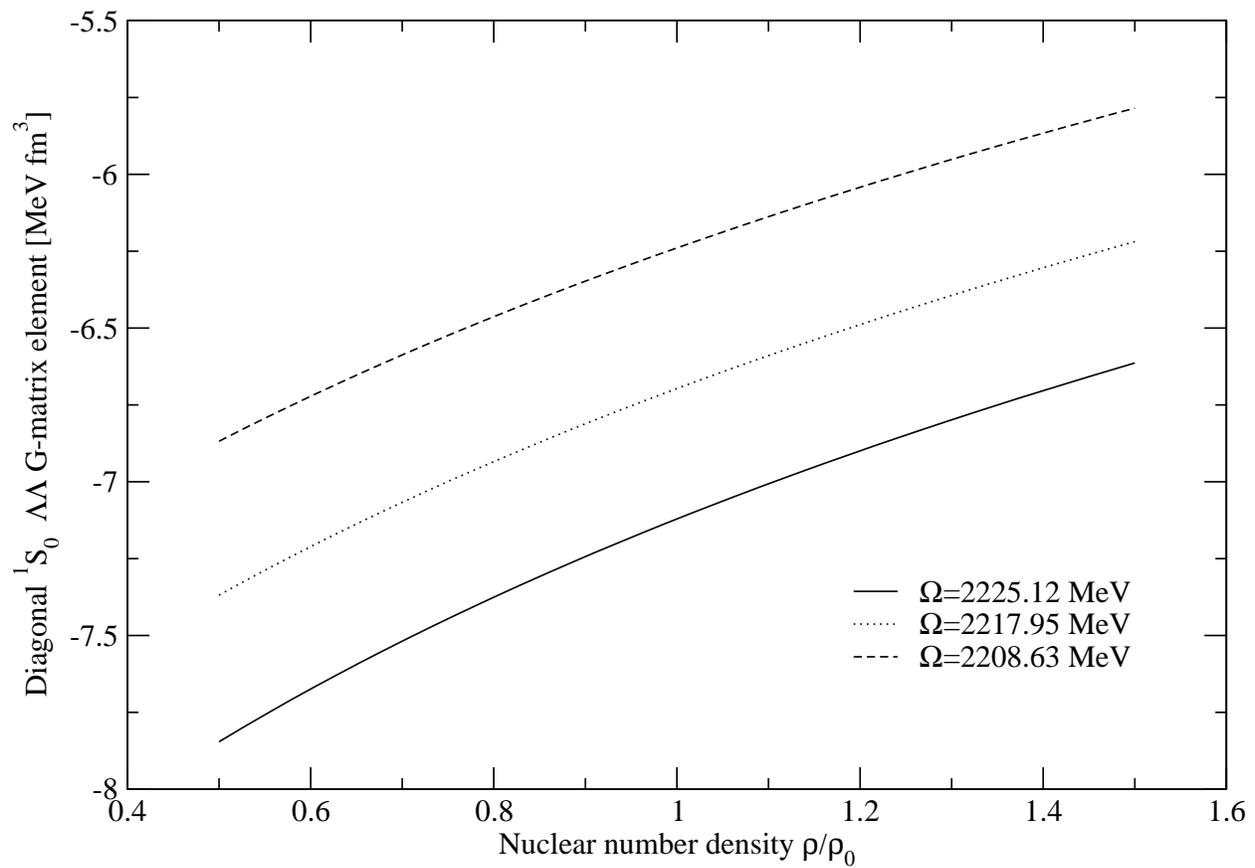}}
      \vspace{0.5cm}
      \caption{$^1S_0$ diagonal $\Lambda\Lambda$ G-matrix element as a function of the nuclear number 
density in units of $\rho_0$, with $\rho_0=0.17$ fm$^{-3}$.}
        \label{fig:gmat}
\end{figure}

The diagonal $^1S_0$ $\Lambda\Lambda$ G-matrix element for zero
center-of-mass momentum and zero relative momentum is shown
in Fig.~\ref{fig:gmat} as a function of the nuclear matter density
for several starting energy values. As density increases, the G-matrix element
loses attraction as a result of Pauli blocking  which reduces the available
phase space for the intermediate $\Xi N$ states. On the other hand, the G-matrix
element gains attraction when the starting energy increases, since the coupling to 
intermediate states is then more efficient.
We will return to this behavior when the results of finite hypernuclei are
discussed.

Table \ref{tab:tab2} displays our results for $\Delta
B_{\Lambda\Lambda}$
in $^{6}_{\Lambda\Lambda}$He, for various coupled-channel cases.
The value of $\Delta B_{\Lambda\Lambda}$ obtained from a calculation
that neglects Pauli blocking effects, {\it i.e.,} directly from the T-matrix, is
also displayed with brackets.
As expected, incorporating the coupling between the $\Lambda\Lambda$ and $\Xi
N$,
produces a drastic effect over the $\Lambda\Lambda$ uncoupled situation,
increasing $\Delta B_{\Lambda\Lambda}$ from 0.16 MeV to 0.78 MeV,
a value that lies very close to the new
experimental datum \cite{Takahashi:nm}. We note that, contrary to what it seems to be implied in 
Ref.~\cite{Afnan:2003ty}, the coupling between the $\Lambda\Lambda$ and $\Xi N$ channels is important even when the 
interaction is weak, as it is the case of the NSC97e potential used here which produces a scattering length of 
about -0.5 fm.
Actually, in Refs.~\cite{Carr:1997ce,Afnan:2003ty} the potential is adjusted for each coupled-channel 
case to reproduce a common value of the scattering length. Therefore, part of the coupling effect is embedded 
in the readjusted parameters.
We now turn to analyzing the effect of the $\Sigma\Sigma$
channel, located more than 150 MeV higher in energy from the
$\Lambda\Lambda$ and $\Xi N$ channels, and which has usually been neglected or taken in an effective way 
within single-channel ($\Lambda\Lambda$) or two-channel ($\Lambda\Lambda,\Xi N$) interaction models. The results
shown in Table \ref{tab:tab2} reveal, surprisingly, that the role of the
$\Sigma\Sigma$ channel is very important and reduces substantially the
two-channel ($\Lambda\Lambda,\Xi N$) value of $\Delta B_{\Lambda\Lambda}$ down to 
0.28 MeV, which is closer to the uncoupled single-channel result. 
Note that the repulsion found for the full coupled-channel G-matrix element around the $\Lambda\Lambda$ threshold does not necessarily mean that the $\Sigma\Sigma$ channel produces a more repulsive interaction. In fact, the $(\Lambda\Lambda,\Xi N,\Sigma\Sigma)$  Nijmegen model becomes so attractive that it even supports a spurious deeply bound YY state around 1500 MeV below the $\Lambda\Lambda$ threshold \cite{yamada}.  However, the 
size of the G-matrix will not be affected by the presence of this bound state since it lies very far away from the region of energies required by our model.
The net effect around the $\Lambda\Lambda$ threshold is that the full coupled-channel calculation has a smaller bond energy than the case in which only the $\Lambda\Lambda$ and the $\Xi N$ channels are retained.
%Actually, a 
%calculation up to second order in perturbation theory would have led to the %opposite 
%result, namely a gain of attraction for the $\Lambda\Lambda$ state and, %therefore, an 
%increase of the bond energy. Our non-perturbative result indicates that the %coupling 
%of the $\Sigma\Sigma$  channel to the $\Lambda\Lambda$ and $\Xi N$ states is %very 
%large, making higher order terms non-negligible and giving rise to this %apparently 
%anomalous decrease of the $\Lambda\Lambda$ bond energy.

Comparing the results of $\Delta B_{\Lambda\Lambda}$ with those between brackets, which have been 
obtained from a T-matrix calculation, one observes that Pauli blocking effects (non-existing in the single
channel $\Lambda\Lambda$ case) are quite important, especially when the
three-channels ($\Lambda\Lambda$, $\Xi N$ and $\Sigma\Sigma$) are considered,
reducing by half the value of $\Delta B_{\Lambda\Lambda}$.
We note that our Pauli unblocked value of 0.54 MeV, obtained for the complete coupled 
channel calculation using the original Nijmegen potential NSC97e, is reasonably close 
to results obtained directly in finite hypernuclei but using effective
Gaussian parameterizations fitted to the scattering length of the Nijmegen NSC97e 
interaction, namely
$\Delta B_{\Lambda\Lambda}=0.58$ MeV \cite{Filikhin:2002wm} and $\Delta 
B_{\Lambda\Lambda}=0.64$ MeV \cite{Myint:2002dp}. We note that the recent variational calculation using the Nijmegen interactions quotes a slightly larger value of 0.81 MeV \cite{yamada}.

The results of Table~\ref{tab:tab2} show that
a proper treatment of Pauli blocking, neglected in most of the calculations using more sophisticated 
ways of treating the finite hypernucleus \cite{hiyama,Albertus:2001pb,Filikhin:2002wm}, is needed to 
draw conclusions on the particular value of $\Delta B_{\Lambda\Lambda}$ predicted by a  given 
interaction. 
A first attempt to incorporate the Pauli suppression effect within the context of finite  
hypernuclei has 
been done recently in Ref.~\cite{Myint:2002dp}, where a Pauli blocking term, correcting 
the phase space of intermediate $\Xi N$ states accessed via
$\Lambda\Lambda \to \Xi N$  conversion up to second order in the effective
interaction, is added. The 
$\Lambda\Lambda$ bond energy is then reduced from 0.64 MeV to 0.21 MeV, hence finding 
a  Pauli suppression of 0.43 MeV, which is about twice the size of the reduction we  
find in the present work, namely $(0.54 -0.28)$ MeV $=0.26$ MeV. The
reason for the difference has to be found in higher order terms of the Pauli 
correction. Indeed, if, in the spirit of the procedure followed in
Ref.~\cite{Myint:2002dp}, we truncate the series that 
defines the $T$-matrix in terms of the $G$-matrix, $T=G+G(1/E-Q/E)T$, up to second order in $G$, 
then the contribution of the Pauli 
blocking correcting term,
$G(1/E-Q/E)G$, amounts to 0.36 MeV. This is consistent with the value 
of 0.43 MeV quoted in Ref.~\cite{Myint:2002dp} which was obtained with a slightly modified
effective interaction to fit the new $\Delta B_{\Lambda\Lambda}$ value in
$^{6}_{\Lambda\Lambda}$He. Moreover, we have checked that the series converges to our T-matrix
result and, hence, to our complete Pauli correction of 0.26 MeV.
The Pauli correction built directly in the finite nucleus in the full coupled-channel calculation of Ref.~\cite{yamada} is also small and of the order of 0.2 MeV.

The scattering length for each coupled-channel situation is also
shown in Table~\ref{tab:tab2} to illustrate, as in other works \cite{Filikhin:2002wm,Afnan:2003ty}, its correlation 
with the $\Lambda\Lambda$ bond energy, which increases as the magnitude of the scattering length increases. This 
correlation is to be expected since $a_{\Lambda\Lambda}$ is proportional to the T-matrix and $\Delta 
B_{\Lambda\Lambda}$ is proportional to the corresponding medium modified G-matrix.
We would also like to point out that the scattering length changes substantially for each of the 
coupled-channel cases. This is apparently different from the results shown in Refs.~\cite{Carr:1997ce,Afnan:2003ty} 
but, as mentioned before, in these later works the interaction is readjusted in each coupled-channel calculation 
to reproduce a common value of the scattering length.

Finally, we collect in Table~\ref{tab:tab3} the results of
$\Delta B_{\Lambda\Lambda}$ for the three observed
$\Lambda\Lambda$ hypernuclei.
The role of coupled channels is qualitatively similar in the three hypernuclei: coupling the $\Xi N$ channel to the 
$\Lambda\Lambda$ channel increases $\Delta B_{\Lambda\Lambda}$ substantially, while the additional incorporation of 
the $\Sigma\Sigma$ channel reduces the binding also substantially, bringing the value of the $\Lambda\Lambda$ 
bond energy closer to the uncoupled result. We also observe that the heavier the nucleus the smaller
the binding, contrary to what one would be expecting from the present experimental results. The trend found 
here is a reflection of the behavior of the $^1S_0$ G-matrix element shown in Fig.~\ref{fig:gmat}. Inspecting the
nuclear structure parameters for each hypernucleus shown in Table~\ref{tab:tab1}, we see that the nuclear 
density for $^{6}_{\Lambda\Lambda}$He is the largest, slightly above 1.5$\rho_0$, hence this hypernucleus has 
the strongest Pauli repulsive effect. However, the starting energy $\Omega$ is also the largest, which produces 
a gain in attraction. For the range of densities and starting energies explored by the three $\Lambda\Lambda$ 
hypernuclear systems studied here, the dependence of the G-matrix on the starting energy is twice more important 
than that on the density. The net effect is that the largest $\Delta B_{\Lambda\Lambda}$ value is obtained for 
the lightest system.

%%%%%%%%%%%%%%%%%%%%%%%%%%%%%%%%%%%%%%%%%%%%%%%%%%%%%%%%%%%%%%%%%%%%%%%%%%%%%%%%%%%%%%%%%%%%%%%%

\section{Conclusions}
\label{sec:conclusions}

In this work we have obtained the bond energy 
$\Delta B_{\Lambda \Lambda}$ in several
$\Lambda\Lambda$ hypernuclei, following a microscopic approach
based on a G-matrix calculation in nuclear matter using,
as $S=-2$ interaction, the recent parameterization NSC97e of the Nijmegen group. We have identified the 
$\Lambda\Lambda$ bond energy with the $^1S_0$ $\Lambda\Lambda$ G-matrix 
element calculated for values of the nuclear density and starting energy appropriate for each hypernucleus.
 
Our simplified finite-nucleus treatment has allowed us to explore in depth
the effect of the various coupled channels and the importance of 
Pauli blocking on the intermediate $\Xi N$ states, paying a special attention 
to the role of the $\Sigma\Sigma$ channel usually neglected in the literature.
Consistently with other works, we find that the coupling between the $\Xi N$ and
$\Lambda\Lambda$ channels has a drastic effect, increasing by about 0.6 MeV the calculated 
$\Delta B_{\Lambda \Lambda}$ in $^{6}_{\Lambda\Lambda}$He with respect to a single-channel 
$\Lambda\Lambda$ calculation. Surprisingly, the additional incorporation of the $\Sigma\Sigma$ channel 
yields a non-negligible reduction in the binding of 0.4 MeV. It would be interesting to explore the role of the coupling to $\Sigma\Sigma$ states in other
three-channel $S=-2$ interactions, such as the Nijmegen hard-core potential F \cite{NijF}. Unfortunately, 
our momentum-space method can only handle soft-core interaction models.
 
We have also explored, within the complete three-channel approach,
the effect of Pauli blocking, which is often neglected or considered in a truncated way in previous works. 
With respect to a T-matrix calculation, our calculated value of $\Delta B_{\Lambda \Lambda}$ 
in $^{6}_{\Lambda\Lambda}$He gets reduced by 0.26 MeV, about half of what was found
on the basis of a second order Pauli corrected calculation \cite{Myint:2002dp}.

Due to our simplified treatment of nuclear structure, we do not expect a quantitative agreement with experimental 
data for the three hypernuclei studied. However, from the bulk of studies of double-$\Lambda$ hypernuclei available 
in the literature, it seems unreasonable to think that a proper finite nucleus calculation which incorporates 
consistently core-polarization effects, might change the calculated bond energies substantially enough to obtain a 
simultaneous agreement with the data. In this respect, our results confirm, in accordance with recent cluster 
calculations \cite{Filikhin:2002wm}, the incompatibility between the experimental binding energies of the light 
double-$\Lambda$ hypernuclear species. We note, however, that the disagreement would be reduced if the $\pi^-$ weak 
decay of the $^{10}_{\Lambda\Lambda}$Be  ground state was assumed to occur to the first excited state of 
$^{9}_{\Lambda}$Be, as pointed out by Filikhin and Gal \cite{Filikhin:2002wm}, hence reducing the bond 
energy in $^{10}_{\Lambda\Lambda}$Be to about 1 MeV. A clarification of the experimental situation, through new 
experiments and analyses, is certainly needed in order to test the theoretical models and make progress in the field 
of doubly-strange systems.  

%%%%%%%%%%%%%%%%%%%%%%%%%%%%%%%%%%%%%%%%%%%%%%%%%%%%%%%%%%%%%%%%%%%%%%%%%%%%%%%%%%%%%%%%%%%%%%%%

\section*{Acknowledgments}

This work is partially supported by DGICYT
project BFM2002-01868 and by the Generalitat de Catalunya project
2001SGR00064.

%%%%%%%%%%%%%%%%%%%%%%%%%%%%%%%%%%%%%%%%%%%%%%%%%%%%%%%%%%%%%%%%%%%%%%%%%%%%%%%%%%%%%%%%%%%%%%%%

%%%%%%%%%%%%%%%%%%%%%%%%%%%%%%%%%%%%%%%%%%%%%%%%%%%%%%%%%%%%%%%%%%%%%%%%%%%%%%%%%%%%%%%%%%%%%%%%%%%%%%%%%%%%%

%%%%%%%%%%%%% FIGURE 1 %%%%%%%%%%%%%%%%%%%%%%%%%%%%%%%%%%
%%%%%%%%%%%%%%%%%%%%%%%%%%%%%%%%%%%%%%%%%%%%%%%%%%%%%%%%%%%

%%%%%%%%%%%%%%%%%%%%%%%%%%%%%%%%%%%%%%%%%%%%%%%%%%%%%%%%%%%%%%%%%%%%%%%%%%%%%%%%%%%%%%%%%%%%%%%%%%%%%%%%%%%%%%

%%%%%%%%%%%%%%%%%%%%%%%%%%%%%%%%%%%%%%%%%%%%%%%%%%%%%%%%%%%%%%%%%%%%%%%%%%%%%%%%%%
\begin{center}
\begin{minipage}{10cm}
\begin{table}
\caption{Parameters of the Woods-Saxon $\Lambda$-nucleus potential ($V_\Lambda$,$a_\Lambda$,
$R_\Lambda$), equivalent $\Lambda$ oscillator parameter ($b_\Lambda$), effective nuclear
density ($\rho$) and G-matrix starting energy ($\Omega$) for each $\Lambda\Lambda$
hypernucleus.}
\vspace{0.5cm}
\begin{tabular}{ c | c | c |c  }
~ & $^{6}_{\Lambda\Lambda}$He \phantom{aaa} & $^{10}_{\Lambda\Lambda}$Be \phantom{aaa} &
$^{13}_{\Lambda\Lambda}$B \phantom{aaa} \\
\hline
$V_\Lambda$ [MeV] & 28 \phantom{aaa} & 28 \phantom{aaa} & 28 \phantom{aaa} \\
$a_\Lambda$ [fm]    & 0.59 \phantom{aaa} & 0.59 \phantom{aaa}  & 0.59 \phantom{aaa} \\
$R_\Lambda$ [fm]    & 1.60 \phantom{aaa} & 2.06 \phantom{aaa}  & 2.67 \phantom{aaa} \\
$b_\Lambda$ [fm] & 2.23 \phantom{aaa} & 1.89 \phantom{aaa} & 1.82 \phantom{aaa} \\
$\rho$ [fm$^{-3}$] & 0.277 \phantom{aaa} & 0.181 \phantom{aaa} & 0.176 \phantom{aaa} \\
$\Omega$ [MeV] & 2224.12 \phantom{aaa} & 2213.67 \phantom{aaa} & 2204.17 \phantom{aaa}
\label{tab:tab1}
\end{tabular}
\end{table}
\end{minipage}
\end{center}
%%%%%%%%%%%%%%%%%%%%%%%%%%%%%%%%%%%%%%%%%%%%%%%%%%%%%%%%%%%%%%%%%%%%%%%%%%%%%%%%%%%%%%%%%%%%%

%%%%%%%%%%%%%%%%%%%%%%%%%%%%%%%%%%%%%%%%%%%%%%%%%%%%%%%%%%%%%%%%%%%%%%%%%%%%%%%%%%%%%%%%%%%%%%

\begin{center}
\begin{minipage}{10cm}
\begin{table}
\caption{$\Lambda\Lambda$ scattering length and $\Lambda\Lambda$ bond energy in $^{6}_{\Lambda\Lambda}$He,
for various channel couplings. Results within brackets ignore Pauli blocking effects.}
\vspace{0.5cm}
\begin{tabular}{c |c |c }
 &  $a_{\Lambda\Lambda}$ [fm] \phantom{aaa} & $\Delta B_{\Lambda\Lambda}$ [MeV] \phantom{aaa} \\
\hline
$\Lambda\Lambda$ & $-0.25$ \phantom{aaa}  & 0.16 (0.16) \phantom{aaa}  \\
$\Lambda\Lambda,\Xi N$ & $-0.84$ \phantom{aaa} & 0.78 (1.02) \phantom{aaa} \\
$\Lambda\Lambda, \Xi N,\Sigma\Sigma$ & $-0.49$ \phantom{aaa} & 0.28 (0.54) \phantom{aaa}
\label{tab:tab2}
\end{tabular}
\end{table}
\end{minipage}
\end{center}

%%%%%%%%%%%%%%%%%%%%%%%%%%%%%%%%%%%%%%%%%%%%%%%%%%%%%%%%%%%%%%%%%%%%%%%%%%%%%%%%%%%%%%%%%%%%%%%%

%%%%%%%%%%%%%%%%%%%%%%%%%%%%%%%%%%%%%%%%%%%%%%%%%%%%%%%%%%%%%%%%%%%%%%%%%%%%%%%%%%%%%%%%%%%%%%%%
\begin{center}
\begin{minipage}{10cm}
\begin{table}
\caption{$\Lambda\Lambda$ bond energy in $^{6}_{\Lambda\Lambda}$He,
$^{10}_{\Lambda\Lambda}$Be and $^{13}_{\Lambda\Lambda}$B, for various channel
 couplings. Units are in MeV.}
\vspace{0.5cm}
\begin{tabular}{c |c |c |c }
 & $^{6}_{\Lambda\Lambda}$He & $^{10}_{\Lambda\Lambda}$Be & $^{13}_{\Lambda\Lambda}$B \\
\hline
$\Lambda\Lambda$ & 0.16 & 0.0046 & 0.11 \\
$\Lambda\Lambda,\Xi N$ & 0.78 & 0.97 & 0.96 \\
$\Lambda\Lambda, \Xi N,\Sigma\Sigma$ & 0.28 & 0.22 & 0.11 \\
\hline
EXP: & $1.01^{+0.38}_{-0.31}$ \cite{Takahashi:nm} &
$4.2 \pm 0.4$  \cite{danysz}
 & $4.8 \pm 0.7$ \cite{aoki}
\label{tab:tab3}
\end{tabular}
\end{table}
\end{minipage}

\end{center}
%%%%%%%%%%%%%%%%%%%%%%%%%%%%%%%%%%%%%%%%%%%%%%%%%%%%%%%%%%%%%%%%%%%%%%%%%%%%%%%%%%%%%%%%%%%%%%%%

\end{document}